\documentclass[twocolumn,amsmath,amssymb,nofootinbib]{revtex4}
\usepackage{amsmath,hyperref}
\usepackage{natbib}
\def\be{\begin{equation}}
\def\ee{\end{equation}}
\def\ba{\begin{eqnarray}}
\def\ea{\end{eqnarray}}
\def\bdm{\begin{displaymath}}
\def\edm{\end{displaymath}}
\def\bwt{\begin{widetext}}
\def\ewt{\end{widetext}}
\def\be{\begin{equation}}
\def\ee{\end{equation}}
\def\ba{\begin{eqnarray}}
\def\ea{\end{eqnarray}}

\def\bdm{\begin{displaymath}}
\def\edm{\end{displaymath}}

\def\bq{\begin{quote}}
\def\eq{\end{quote}}

 at 10truept

\newcommand{\beq}{\begin{equation}}
\newcommand{\eeq}{\end{equation}}
\newcommand{\beqa}{\begin{eqnarray}}
\newcommand{\eeqa}{\end{eqnarray}}
\newcommand{\mpl}{M_{Pl}}

\begin{document}

\title{Reissner-Nordstr\"{o}m Black Holes on a Codimension-2 Brane}

\author{Derrick Kiley}
\email{kileydt@lacitycollege.edu}
\affiliation{Department of Physics and Engineering, Los Angeles City College, Los Angeles,
CA 90029}

\date{\today}

\begin{abstract}
Here we derive the exact Reissner-Nordstr\"{o}m black hole solution on a tensional codimension-2 brane, generalizing earlier Schwarzschild and Kerr results.  We begin by briefly reviewing various aspects of codimension-2 branes that will be important for our analysis, including the mechanism of ``offloading'' of brane tension into the bulk that is unique to these branes, as well as the explicit construction of the codimension-2 Schwarzschild black hole as a warm-up exercise.  We then show that the same methods can be used to find the metric describing the spacetime surrounding an electrically-charged point source threaded by a codimension-2 brane.  The presence of the brane tension leads to an amplification of the apparent strength of gravity, as is well-known, and we further find exactly the same enhancement for the apparent strength of the electric field.

\end{abstract}

\maketitle

\section{Introduction}

In the area of gravitational research, few topics have received more attention than black holes, which nevertheless remain mysterious.  Ideas from string theory add another layer of mystery, positing that the Universe could have additional hidden dimensions.  Because gravitation can be viewed as the curvature of space-time, gravitational sources should make their effects felt throughout all of these dimensions.  Since gravity would thus be a fundamentally higher-dimensional phenomenon, only appearing four-dimensional to our coarse senses, the intrinsic strength of gravity can be different than the \emph{effective} four-dimensional value we observe.  This difference between the fundamental and observed strengths can potentially provide a ``solution'' to the hierarchy problem \cite{ArkaniHamed:1998rs} - \cite{Dvali:2000hr} (although the problem is then shifted to explaining why the extra dimensions are compactified and in what form).

String theory further predicts higher-dimensional surfaces (\emph{branes}) that can live in the extra dimensional space (the \emph{bulk}).  A particularly interesting brane configuration is a codimension-2 brane, in which a $D$-dimensional brane floats in a $D+2$-dimensional bulk; for example, a three brane (one time and three space dimensions) living in a six-dimensional space-time.  These branes have a unique and useful property.  Typically the presence of matter on a brane can curve the brane, as well as the surrounding bulk.  However, in the case of a codimension-2 brane, vacuum energy (called the brane \emph{tension}) can be ``offloaded'' into the bulk, leaving the brane flat.  The bulk then acquires the topology of a cone with the tip centered on the brane.  The bulk remains locally flat, but gains a conical singularity along the brane in a way that is similar to that in the space surrounding a cosmic string, as is well-known \cite{Aryal:1986sz}.  Let us briefly review how this happens \cite{Kaloper:2006ek}.

We begin with the action describing six-dimensional bulk gravity and include a three-brane with matter Lagrangian $\mathcal{L}_4$,
\be
S=\frac{M_6^4}{2}\int d^6x \sqrt{-g_6} R_6 + \int d^4x \sqrt{-g_4} \mathcal{L}_4,\\
\label{eqn:gravaction}
\ee
where $M_6$ is the \emph{fundamental} six-dimensional Planck scale (in units $c=\hbar \equiv 1$).  Variation of the action in Eq. (\ref{eqn:gravaction}) yields the six-dimensional Einstein equations
\be
M_6^4 G^M{}_N = T^\mu{}_\nu \delta^M{}_\mu \delta^\nu{}_N \frac{\delta^{(2)}\left(\vec{y}\right)}{\sqrt{h}}.\\
\label{eqn:6DEinstein}
\ee
Here $T^\mu{}_\nu$ is the four-dimensional brane stress energy tensor localized to the brane by the delta function.  The indices $M,N$ run from 0 to 5, while $\mu,\nu$ run from 0 to 3, and $\vec{y}$ denotes the extra two bulk coordinates.  Here $h$ is the determinant of the bulk components of the metric, $\sqrt{h} = \sqrt{-g_6}/\sqrt{-g_4}$.  Tracing Eq. (\ref{eqn:6DEinstein}) gives the 6D Ricci scalar
\be
R_6 = -\frac{T}{2M_6^4} \frac{\delta^{(2)}\left(\vec{y}\right)}{\sqrt{h}},\\
\label{eqn:6DRicciSch}
\ee
where $T \equiv T^\mu{}_\mu$ is the trace of the brane stress-energy tensor.  We can now insert Eq. (\ref{eqn:6DRicciSch}) back into Eq. (\ref{eqn:6DEinstein}) to find
\be
R^M{}_N = \frac{1}{M_6^4}\left(T^\mu{}_\nu \delta^M{}_\mu \delta^\nu{}_N - \frac{1}{4}\delta^M{}_N T\right) \frac{\delta^{(2)}\left(\vec{y}\right)}{\sqrt{h}}.\\
\label{eqn:6DRiccitensorSch}
\ee
Eq. (\ref{eqn:6DRiccitensorSch}) can be broken up into bulk and brane pieces (here $a$ and $b$ denote the two bulk coordinates),
\be
R^\mu{}_\nu= \frac{1}{M_6^4}\left(T^\mu{}_\nu - \frac{1}{4}\delta^\mu{}_\nu T\right) \frac{\delta^{(2)}\left(\vec{y}\right)}{\sqrt{h}},\\
\label{eqn:Riccibrane}
\ee
\be
R^a{}_b = -\frac{T}{4M_6^4}\delta^a{}_b  \frac{\delta^{(2)}\left(\vec{y}\right)}{\sqrt{h}}.
\label{eqn:Riccibulk}
\ee
We now split up the brane stress-energy tensor into a tensional contribution with brane tension $\lambda$, and a piece denoting any other matter contributions, $\tau^\mu{}_\nu$,
\be
T^\mu{}_\nu = - \lambda \delta^\mu{}_\nu + \tau^\mu{}_\nu.\\
\label{eqn:4DSEtensor}
\ee
Plugging in Eq. (\ref{eqn:4DSEtensor}) to Eq. (\ref{eqn:Riccibrane}) gives our final result along the brane,
\be
R^\mu{}_\nu = \frac{1}{M_6^4}\left[\tau^\mu{}_\nu - \frac{1}{4}\delta^\mu{}_\nu\tau\right] \frac{\delta^{(2)}\left(\vec{y}\right)}{\sqrt{h}},\\
\label{eqn:Riccibranetau}
\ee
where $\tau \equiv \tau^\mu{}_\mu$ is the trace of the non-tensional stress-energy tensor on the brane.  In the bulk we plug in Eq. (\ref{eqn:4DSEtensor}) to Eq. (\ref{eqn:Riccibulk}) to find
\be
R^a{}_b = \frac{1}{M_6^4}\left[\lambda - \frac{1}{4}\tau\right]\delta^a{}_b \frac{\delta^{(2)}\left(\vec{y}\right)}{\sqrt{h}}.\\
\label{eqn:Riccibulktau}
\ee

In Eqs. (\ref{eqn:Riccibranetau}) and (\ref{eqn:Riccibulktau}) we see a remarkable result: the brane tension has vanished from the brane equations, but does appear in the bulk equations.  This is the ``offloading'' of the tension into the bulk, as discussed earlier.  We now suppose that the brane contains only tension, such that $\tau^\mu{}_\nu =0$.  Then Eq. (\ref{eqn:Riccibranetau}) reduces to $R^\mu{}_\nu=0$, while Eq. (\ref{eqn:Riccibulktau}) becomes
\be
R^a{}_b = \frac{\lambda}{M_6^4}\delta^a{}_b \frac{\delta^{(2)}\left(\vec{y}\right)}{\sqrt{h}}.\\
\label{eqn:Riccibulklambda}
\ee
Thus, we see that the brane Ricci tensor vanishes, while the bulk Ricci terms contain a delta function spike leading to a conical singularity, as we will now demonstrate.

Because the brane Ricci tensor vanishes  the Ricci scalar is simply the trace of the two-dimensional Ricci tensor in Eq. (\ref{eqn:Riccibulklambda}), $R_6 = R_2$ where
\be
R_2 = \frac{2\lambda}{M_6^4}\frac{\delta^{(2)}\left(\vec{y}\right)}{\sqrt{h}}.\\
\label{eqn:2DRicciscalar}
\ee
We take for the two-dimensional bulk metric the \emph{ansatz} $g_{ab} = e^{-2\theta}\delta_{ab}$, where $\theta$ is a function of $\vec{y}$, such that the bulk coordinates are conformally flat.  In this case the Ricci tensor is $R_2 = 2e^{2\theta}\vec{\nabla}_{\vec{y}}^2\theta$.  Plugging in the metric \emph{ansatz} to Eq. (\ref{eqn:2DRicciscalar}) and using $\sqrt{h} = \sqrt{\det\left|e^{-2\theta}\delta_{ab}\right|} = e^{-2\theta}$ we find for Eq. (\ref{eqn:2DRicciscalar}) $R_2 = \left(2\lambda/M_6^4\right) e^{2\theta}\delta^{(2)}\left(\vec{y}\right)$.  Comparing both expressions for $R_2$ we see that
\be
\vec{\nabla}_{\vec{y}}^2\theta = \frac{\lambda}{M_6^4}\delta^{(2)}\left(\vec{y}\right).\\
\label{eqn:Laptheta}
\ee
Thus, $\theta$ is solved by the two-dimensional Green's function.  Noting that $\vec{\nabla}_{\vec{y}}^2\ln\left(|\vec{y}|/\ell\right) = 2\pi \delta^{(2)}\left(\vec{y}\right)$, where $\ell$ is a scaling constant needed to make the units work out, we finally find that $\theta = 2b \ln\left(|\vec{y}|/\ell\right)$ where
\be
b = \frac{\lambda}{2\pi M_6^4}.\\
\label{eqn:b}
\ee
Hence $g_{ab}dx^a dx^b = \left(|\vec{y}|/\ell\right)^{-2b}\left(dy_1^2+dy_2^2\right)$.  By making the coordinate transformations
\be
\begin{array}{ccl}
y_1 & = & \left[\left(1-b\right)\frac{\rho}{\ell^b}\right]^{1/(1-b)}\cos\phi\\
y_2 & = & \left[\left(1-b\right)\frac{\rho}{\ell^b}\right]^{1/(1-b)}\sin\phi,\\
\end{array}
\label{eqn:rhotrans}
\ee
the metric becomes $ds_2^2 = d\rho^2 + \left(1-b\right)^2 \rho^2 d\phi^2$, which is \emph{almost} flat polar coordinates, except for the factor of $\left(1-b\right)^2$ in front of the $d\phi^2$ term.  This shows the conical singularity; the polar angle $\phi$ is rescaled, $\phi \rightarrow \left(1-b\right)\phi$ meaning that it does not run around a full $2\pi$ radians, but rather from $0$ to $\left(1-b\right)\times 2\pi$ (there is a deficit angle).  We see that the net effect of the brane tension is to extract a wedge of angle $\delta \equiv 2\pi b = \lambda/M_6^4$ from the flat bulk space.  The edges of the wedge cut are then identified, yielding the conical space.  One can also check that this metric represents a bulk conical space by embedding a cone into a \emph{seven}-dimensional Minkowski space and checking that the resulting $6D$ metric comes out the same \cite{Kiley:2008zz}.

Now, in the absence of any non-tensional matter whatsoever the full six-dimensional metric may be written $ds_6^2 = \eta_{\mu\nu}dx^\mu dx^\nu + d\rho^2 + \left(1-b\right)^2 \rho^2 d\phi^2$.  Let us rewrite $ds_6^2$ in spherical coordinates.  Noting that $\eta_{MN}dx^M dx^N = -dt^2 + dr^2 + r^2 d\Omega_4^2$, we can include the effect of the conical deficit by writing $d\Omega_4^2 = d\Omega_3^2 + \left(1-b\right)^2\Pi_{k=1}^3\sin^2\left(\theta_k\right)d\chi^2$ \cite{Kaloper:2006ek}.  Thus, the flat metric, including the brane tension is simply
\be
\begin{array}{ccl}
ds_6^2 & = & - dt^2 + dr^2 +r^2\left\lbrace d\theta^2+ \sin^2\theta \left[d\phi^2 \right.\right.\\
&&\left.\left.\quad \quad+ \sin^2 \phi \left(d\psi^2+\left(1-b\right)^2 \sin^2\psi \,d\chi^2\right)\right]\right\rbrace,\\
\end{array}
\label{eqn:flatconicalspace}
\ee
which is flat space, but with a conical deficit.  The metric in Eq. (\ref{eqn:flatconicalspace}) explicitly contains the conical singularity and solves $R_{MN} = R_6 =0$ away from the brane.  The metric makes sense for all values of the brane tension, including for supercritical values when $b>1$.  In this case the bulk spacetime is compactified into a two-dimensional teardrop shape \cite{Kaloper:2007ap}.

Because the brane tension is offloaded into the bulk, finding gravitational solutions is considerably simplified.  On the other hand, it appears that the presence of non-tensional stress energy on the brane would ruin this simple offloading because of Eq. (\ref{eqn:Riccibulktau}), which includes the matter contributions.  However, it is obvious that allowing matter with a vanishing trace of its stress-energy tensor (such as relativistic matter) to live on the brane would keep the bulk geometry simple, even if the brane geometry becomes complicated.  Furthermore, as we will see below, the bulk can remain conical, even in the presence of some further brane sources.

A particularly important system with vanishing stress-energy is a black hole, which satisfies the vacuum Einstein equations away from the singularity.  Generalizing the black hole solutions to higher-dimensions is straightforward for an empty bulk \cite{Myers:1986un} - \cite{Myers:2011yc}, but exact black hole solutions including a brane are typically difficult to find.  Fortunately, in the codimension-2 case the solution is very straightforward and the \emph{exact} Schwarzschild solution, including a tensional brane, was constructed in \cite{Kaloper:2006ek}.  While the Schwarzschild black hole solution is completely characterized by its mass, other black hole solutions exist involving spin and electric (and magnetic) charge.  The solution describing a rotating black hole on a codimension-2 brane was made in \cite{Kiley:2007wb}, while the solution including electric charge, the Reissner-Nordstr\"{o}m solution, is the subject of the remainder of this paper.

\section{Reviewing the Schwarzschild Case}

Before turning to the Reissner-Nordstr\"{o}m solution, let us \emph{very} briefly review the much simpler Schwarzschild case, originally constructed in \cite{Kaloper:2006ek}, and explicitly determine the exact six-dimensional solution.  The Reissner-Nordstr\"{o}m solution will proceed along the same lines, and so the aside will be well worth it.  

We have seen that the conical/Minkowski space in Eq. (\ref{eqn:flatconicalspace}) allows for vanishing Ricci terms off the bulk, but the Minkowski metric is not the only solution of $R = 0$, as is very well-known.  Let us instead try a metric of the form
\be
\begin{array}{ccl}
ds_6^2 & = & - f(r)dt^2 + \frac{dr^2}{f(r)} +r^2\left\lbrace d\theta^2+ \sin^2\theta \left[d\phi^2 \right.\right.\\
&&\left.\left.\quad \quad + \sin^2 \phi \left(d\psi^2+\left(1-b\right)^2 \sin^2\psi \,d\chi^2\right)\right]\right\rbrace,\\
\end{array}
\label{eqn:f(r)conicalspace}
\ee
where $f$ is a function of only the radial coordinate in order to maintain isotropy of the space-time, and we explicitly include the brane tension.  Working out the Ricci tensor components gives
\be
\begin{array}{ccl}
R_{tt} & = & \frac{1}{2}\left(\frac{d^2f}{dr^2}+\frac{4}{r}\frac{df}{dr}\right)g_{tt}\\
R_{rr} & = & -\frac{1}{2}\left(\frac{d^2f}{dr^2}+\frac{4}{r}\frac{df}{dr}\right)g_{rr}\\
R_{mn} & = & -\frac{1}{r^2}\left(r \frac{df}{dr}+3f-3\right)g_{mn},\\\end{array}\\
\label{eqn:f(r)Ricci}
\ee
where $m$ and $n$ stand for the remaining coordinates, $\lbrace \theta, \phi, \psi, \chi\rbrace$.  Every term in Eqs. (\ref{eqn:f(r)Ricci}) is independent of the brane tension, except for the contribution from $g_{\chi\chi}$.  Solving for the Ricci scalar also gives a result independent of the brane tension,
\bdm
R_6 = -12 + 12 f(r) + 8 r \frac{df}{dr} + r^2 \frac{d^2f}{dr^2} .\\
\edm
Choosing $R_{mn} = 0$ sets $rf'+3f-3=0$, such that
\bdm
f(r) = 1 + \frac{C}{r^3},\\
\edm
where $C$ is an integration constant (this solution for $f$ also satisfies all of the other equations, $R_{MN} = R_6 \equiv 0$ as expected).  We can determine the constant $C$ as follows: we write $g_{tt} = g^0_{tt} + h_{tt} = -\left(1 - h_{tt}\right) = -\left(1+C r^{-3}\right)$, setting $h_{tt} = -Cr^{-3}$.  However, the ADM mass is \cite{Myers:2011yc}
\bdm
h_{tt} \approx \frac{1}{8\Omega_4\left(b\right)M_6^4}\frac{M}{r^3},\\
\edm
where $\Omega_4\left(b\right)$ is the four-dimensional surface area \emph{including the brane tension}, and $M$ is the mass.  Explicitly, $\Omega_4 = \frac{8\pi^2}{3}\left(1-b\right)$, and so comparing the results gives
\bdm
C = -\frac{3}{64\pi^2 \left(1-b\right)M_6^4}M.\\
\edm
Note the appearance of the brane tension parameter, $b$ in this expression.  Thus, the generalization of the six-dimensional Schwarzschild metric that is threaded by a tensional brane is Eq. (\ref{eqn:f(r)conicalspace}) with
\be
f(r) = 1-\frac{3}{64\pi^2\left(1-b\right)M_6^4}\frac{M}{r^3}.\\
\label{eqn:f(r)Schfull}
\ee

This expression differs from the ordinary result in six dimensions \cite{Myers:2011yc} by the addition of the $(1-b)^{-1}$ factor.  The metric result in Eqs. (\ref{eqn:f(r)conicalspace}) and (\ref{eqn:f(r)Schfull}) could have easily been found by starting with the ordinary six-dimensional Schwarzschild solution, and then extracting a wedge of deficit angle $\delta = 2\pi b$ from the bulk coordinates, arriving at the solution immediately.  The solution was first constructed in \cite{Kaloper:2006ek}, in just this way.

The presence of the brane tension in Eq. (\ref{eqn:f(r)Schfull}) leads to some interesting effects.  In particular, the event horizon $r_H$ is found from setting $f(r)=0$,
\be
r_H = \left(\frac{3M}{64\pi^2\left(1-b\right)M_6^4}\right)^{1/3} \equiv \frac{r_0}{\left(1-b\right)^{1/3}},\\
\label{eqn:Schhorizon}
\ee
where $r_0$ is the ordinary result in six dimensions, if the brane was not present.  The larger the brane tension becomes, such that $b \rightarrow 1$, then the larger the horizon size grows.  This effect, dubbed the ``lightning rod effect,'' \cite{Kaloper:2006ek} is conceptually easy to understand; the gravitational field lines from the black hole are confined to the surface of a cone in the bulk dimensions.  This keeps the field lines together such that they cannot spread out and dilute as quickly as they would in an otherwise flat space, much like the electric field around a needle can grow very large even though there may only be a small amount of charge.

Because the gravity does not dilute as quickly, this enhancement of the event horizon can be viewed in another way.  Gravity appears stronger than would be expected based on a naive analysis including the fundamental gravitational scale, $M_6$.  However, we can define an \emph{effective} six-dimensional gravitational scale
\be
M_{6\textrm{eff}}^4 \equiv \left(1-b\right)M_6^4,\\
\label{eqn:effMpl}
\ee
which includes the conical enhancement.  Thus, we see that the net effect of the brane tension is to rescale the gravitational scale, amplifying it.  Upon compactification of the extra dimensions to a scale $\sim L$, then the four-dimensional Plank mass $\mpl^2 \sim L^2 M_{6\textrm{eff}}^4$, and hence the 4D Planck mass that we observe on everyday scales would thus already include any effects from brane tension \cite{Kaloper:2006ek}.

The amplification due to the brane tension can make its effects felt in additional ways, for example increasing the lifetime of an evaporating black hole and changing the angular momentum of a spinning black hole \cite{Kiley:2007wb}.  We will also see below that the amplification is not limited to gravity, but can also affect the electrical field of a charged mass for the same reasons as discussed for gravity: the electrical field lines also cannot dilute as quickly, and so electricity appears stronger.  Now that we have seen the method for the Schwarzschild solution, we generalize to electrically charged black holes.

\section{The Reissner-Nordstr\"{o}m Solution}

While the analysis describing an electrically-charged black hole is performed in the same way as in the Schwarzschild case, the algebra is a bit harder since the Ricci scalar does not vanish in this case.  We begin with the Einstein-Maxwell Lagrangian
\be
\begin{array}{ccl}
S & = &\frac{M_6^4}{2}\int d^6x \sqrt{-g_6} R_6 - \frac{1}{4\mu_6}\int d^6x \sqrt{-g_6}F_{MN}F^{MN}\\
&& + \int d^4x \sqrt{-g_4} \mathcal{L}_4,\\
 \end{array}
\label{eqn:EMgravaction}
\ee
where the second term is the new addition.  Here $F^{MN}$ is the six-dimensional electromagnetic field tensor, and we use slightly uncommon ``SI-like" units, explicitly including the six-dimensional permeability of free space for later convenience.  Variation of the action in Eq. (\ref{eqn:EMgravaction}) with respect to the metric yields the gravitational equations,

\be
\begin{array}{ccl}
R^M{}_N - \frac{1}{2}\delta^M{}_N R & = &  \frac{1}{M_6^4\mu_6}\left[F^{MP}F_{NP}-\frac{1}{4}\delta^M{}_N F^{PQ}F_{PQ}\right]\\
&&\quad +\frac{1}{M_6^4}T^\mu{}_\nu \delta^M{}_\mu \delta^\nu{}_N\frac{\delta^{(2)}\left(\vec{y}\right)}{\sqrt{h}},\\
\end{array}
\label{eqn:EMeqns}
\ee
while variation with respect to the electromagnetic field $A^M$ says that the field tensor must be divergenceless,
\be
\nabla_M F^{MN}=0.\\
\label{eqn:6DMaxwelleqns}
\ee
Once again we can split up Eq. (\ref{eqn:EMeqns}) into bulk and brane pieces, taking the brane stress-energy tensor $T^\mu{}_\nu = - \lambda \delta^\mu{}_\nu$ to include only tension,  

\be
R^a{}_b = - \frac{1}{8M_6^4\mu_6}F_{PQ}F^{PQ}\delta^a{}_b + \frac{\lambda}{M_6^4}\delta^a{}_b \frac{\delta^{(2)}\left(\vec{y}\right)}{\sqrt{h}}.\\
\label{eqn:6DEMbulk}
\ee
\be
R^\mu{}_{\nu} = \frac{1}{M_6^4 \mu_6}\left(F^{\mu P}F_{\nu P} - \frac{1}{8}\delta^\mu{}_\nu F_{PQ}F^{PQ}\right),\\
\label{eqn:6DEMbrane}
\ee
where the tension again cancels from the brane equations.  These are the equations that we need to solve.

Notice that the bulk equations in Eq. (\ref{eqn:6DEMbulk}) contain electromagnetic components, in contrast to Eq. (\ref{eqn:Riccibulklambda}).  At first sight this would seem to complicate matters considerably since the bulk is curved from the electromagnetic field.  However, this curvature is just what one would expect from the ordinary 6D Reissner-Nordstr\"{o}m metric, as we will see.  We begin again with a tensionless metric \emph{ansatz},

\be
ds_6^2 = -f(r)dt^2 + \frac{dr^2}{f(r)} + r^2 d\Omega_4^2.\\
\label{eqn:6DRNnobrane}
\ee
In what follows it will be convenient to briefly make a coordinate transformation, defining a new ``radial'' distance,
\be
\mathcal{R} \equiv \exp\left[K\int \frac{dr}{r\sqrt{f(r)}}\right],\\
\label{eqn:uniformtrans}
\ee
where $K$ is a scaling constant.  This transforms the metric into \emph{uniform coordinates},
\be
ds_6^2 = - F\left(\mathcal{R}\right) dt^2+ G\left(\mathcal{R}\right)\left(d\mathcal{R}^2 +\mathcal{R}^2 d\Omega_4^2\right),\\
\label{eqn:6DRNuniform}
\ee
where $G\left(\mathcal{R}\right) = r^2\left(\mathcal{R}\right)\exp\left[-2K\int \frac{dr}{r\sqrt{f}}\right]$, and $F\left(\mathcal{R}\right) = G^{-1}\left(\mathcal{R}\right)\left(\frac{dr}{d\mathcal{R}}\right)^2$ depend directly on the metric function $f(r)$, but their exact forms are not important for this analysis.  In these coordinates the bulk metric components are once again conformally flat
\be
g_{ab} = G\left(\mathcal{R}\right)\delta_{ab}.\\
\label{eqn:old6DRNbulkmetric}
\ee

To include the brane tension, we again try $\tilde{g}_{ab} = e^{-2\theta}g_{ab}$, then $\tilde{R}_{ab} = R_{ab} + \delta_{ab}\vec{\nabla}^2 \theta$, where $R_{ab}$ is the Ricci tensor in the absence of the brane tension.  Noting that
\bdm
\frac{\lambda}{M_6^4}\tilde{g}_{ab}\frac{\delta^{(2)}\left(\vec{y}\right)}{\sqrt{h}} = \frac{\lambda}{M_6^4}\delta_{ab}\delta^{(2)}\left(\vec{y}\right),\\
\edm
which we can set equal to $\delta_{ab} \vec{\nabla}^2\theta$, we once again find Eq. (\ref{eqn:Laptheta})!  Thus, the brane tension affects the electrically-charged black hole in precisely the same way as it does the Schwarzschild black hole, with $b$ once again given by Eq. (\ref{eqn:b}).

Using the same coordinate transformations in Eqs. (\ref{eqn:rhotrans}) gives (upon setting $\mathcal{R}^2 \equiv \vec{x}^2 + \rho^2$, where $\vec{x}^2$ is the three-dimensional brane distance)
\bdm
\tilde{g}_{ab} = G\left(\mathcal{R}\right)\left(d\rho^2 + \left(1-b\right)^2 d\phi^2\right).\\
\edm
This is the same metric as the flat case, except for the conformal factor of $G\left(\mathcal{R}\right)$.  We can now undo the coordinate transformation to go back to the metric form in Eq. (\ref{eqn:6DRNnobrane}), in terms of $f(r)$, which now includes the brane tension parameter, $b$.  This form simply gives back Eq. (\ref{eqn:f(r)conicalspace}) again (although $f$ is different for the electromagnetic case, of course), and the Ricci terms are again given by Eqs. (\ref{eqn:f(r)Ricci}).

We can determine $f(r)$, by solving Eq. (\ref{eqn:6DEMbrane}), but we first need the $F_{PQ}F^{PQ}$, which we can find from Eq. (\ref{eqn:6DMaxwelleqns}).  For our problem, we are only interested in a static, electrically-charged black hole with electric field $E\left(r\right)$.  In this case, $F^{MN} = E(r)\left(\delta^M{}_t\delta^N{}_r - \delta^M{}_r\delta^N{}_t\right)$.  Then, Eq. (\ref{eqn:6DMaxwelleqns}) gives $\partial_r\left(\sqrt{-g_6}E\right) =0$.  From the metric in Eq. (\ref{eqn:6DRNnobrane}) $\sqrt{-g_6} \sim r^4$, such that $ \partial_r\left(r^4 E\right) = 0$.  Hence, $E = \textrm{const}/r^4$.  To determine the constant, we can use the 6D Gauss's law, $\oint \vec{E}\cdot d\vec{A}_4 = Q/\epsilon_6$, where $Q$ is the electric charge, and we again use ``SI-like'' coordinates with a generalized 6D permittivity of free space.  The \emph{effective} four-dimensional  permeability and permittivities are found in terms of the six-dimensional fundamental values via compactification, $\mu_0 \sim \mu_6 L^2$ and $\epsilon_0 \sim \epsilon_6 L^{-2}$ where $L$ is again the compactification radius.  Then the 6D quantities satisfy $\mu_6\epsilon_6 = \mu_0\epsilon_0 =c^{-2} \equiv 1$, in our units.  

Now, solving Gauss's law on a spherical Gaussian surface gives $\oint \vec{E}\cdot d\vec{A}_4 = E \Omega_4\left(b\right)r^4 = Q/\epsilon_6$, where $\Omega_4\left(b\right)$ is once again the four-dimensional surface area including the brane tension.   Thus, we finally find
\be
E\left(r\right) = \frac{3Q}{8\pi^2\left(1-b\right)\epsilon_6 r^4},\\
\label{eqn:6DE(r)}
\ee
which again contains the brane tension.  Then, using Eq. (\ref{eqn:6DE(r)}) we have
\bdm
F_{PQ}F^{PQ} = -\frac{9Q^2}{32\pi^2 (1-b)^2\epsilon_6^2 r^8}.\\
\edm

Once again, the $R_{mn}$ may be found from Eq. (\ref{eqn:6DEMbrane}) and Eqs. (\ref{eqn:f(r)Ricci}), which gives
\bdm
\frac{9Q^2}{32\pi^2\left(1-b\right)^2 M_6^4\epsilon_6 r^8}g_{mn} = - \frac{1}{r^2}\left(r\frac{df}{dr}+3f-3\right)g_{mn},\\
\edm
after setting $\mu_6 \epsilon_6 =1$ in our units.  Solving for $f\left(r\right)$ gives
\be
f(r) = 1 + \frac{C}{r^3}+\frac{3Q^2}{32\pi^2\left(1-b\right)^2M_6^4 \epsilon_6}\frac{1}{r^6},\\
\label{eqn:f(r)almost}
\ee
where $C$ is again an integration constant.  Comparing with the ADM mass (or, more simply noting that Eq. (\ref{eqn:f(r)almost}) must reduce to Eq. (\ref{eqn:f(r)Schfull}) when the electric charge goes to zero) gives same results,
\bdm
C = -\frac{3}{64\pi^2\left(1-b\right)M_6^4}M.\\
\edm
So, the full (and exact)  metric describing an electrically-charged six-dimensional black hole threaded by a tensional brane is Eq. (\ref{eqn:f(r)conicalspace}) with
\be
f(r) = 1 -\frac{3}{64\pi^2\left(1-b\right)M_6^4} \frac{M}{r^3}+\frac{3Q^2}{32\pi^2\left(1-b\right)^2 M_6^4\epsilon_6}\frac{1}{r^6}.\\
\label{eqn:f(r)full}
\ee

Notice in the term $\sim r^{-6}$ \emph{two} factors of $(1-b)^{-1}$ appear, whereas only one factor appears in the term $\sim r^{-3}$.  This is easy to understand; one factor can be absorbed into the Planck scale, $M_{6\textrm{eff}}^2 \sim M_6^2(1-b)$, while the other factor rescales the electric field strength, $\epsilon_{6\textrm{eff}} \sim (1-b) \epsilon_6$, in the same way (this explains why we chose the ``SI-like'' units).  As discussed above, this is to be expected since the electric field lines will also not spread out as quickly in the conical bulk as they could in a fully six-dimensional spacetime.  Thus, both the gravitational and electrical forces experience the lightning rod effect discussed above.  Once again, the horizons are affected by the presence of the brane, in precisely the same way as Eq. (\ref{eqn:Schhorizon}), $r_H =\left(1-b\right)^{-1/3}r_{0\textrm{RN}}$, where $r_{0\textrm{RN}}$ are the ordinary ``braneless'' six-dimensional Reinsser-Nordstr\"{o}m horizons.  This tensional enhancement of the electrical field would also likely lead to a more rapid electrical discharge of the black hole.

\section{Conclusions}

We have explicitly constructed the \emph{exact} metric describing an electrically-charged point mass threaded by a tensional codimension-2 brane in a six-dimensional space-time.  This analysis demonstrates a solution to a codimension-2 brane system where the matter stress-energy tensor does not vanish in the bulk, unlike the previous tensional brane black hole solutions found in \cite{Kaloper:2006ek} and \cite{Kiley:2007wb}.  We have found that the presence of the brane tension enhances the apparent strength of the gravitational and also electrical fields surrounding the black hole.

The previous black hole solutions in \cite{Kaloper:2006ek} and \cite{Kiley:2007wb} could be constructed simply by starting with the empty-bulk six-dimensional solution, extracting a wedge from an axis of symmetry by rescaling $\chi \rightarrow \left(1-b\right)\chi$ and rescaling the fundamental six-dimensional Planck constant.  Here we find that the electrically-charged black hole solutions contain an additional affect owing to the identical tensional enhancement of the electric field. So, in this case, the final metric is a bit more subtle than a simple construction would suggest, although there are no real surprises in the final result, Eq. (\ref{eqn:f(r)full}).

\section{Acknowledgments}

I would like to thank N. Kaloper, for interesting and useful comments.

\end{document}